\documentstyle[aps,prb] {revtex} 
\tighten
\preprint{10 August 1998}

\begin{document}
\draft
 \twocolumn
\title{ Frequency-dependent magnetotransport and particle dynamics 
in magnetic modulation systems}
 
\author{Esmael Badran and Sergio E. Ulloa}
\address{
 Department of Physics and Astronomy and Condensed Matter and Surface
Sciences Program, Ohio University, Athens, OH 45701-2979}

\maketitle 

\begin{abstract} 
 We analyze the dynamics of a charged particle moving in the presence
of spatially-modulated magnetic fields.  From Poincare surfaces of
section and Liapunov exponents for characteristic trajectories we find
that the fraction of pinned and runaway quasiperiodic orbits {\em vs}.
chaotic orbits depends strongly on the ratio of cyclotron radius to the
structure parameters, as well as on the amplitude of the modulated
field.  We present a complete characterization of the dynamical
behavior of such structures, and investigate the contribution to the
magnetoconductivity from all different orbits using a classical Kubo
formula.  Although the DC conductivity of the system depends strongly
on the pinned and runaway trajectories, the frequency response reflects
the topology of all different orbits, and even their unusual temporal
behavior.  \\
 \end{abstract}

\pacs{PACS Nos. 73.20.Dx, 72.10.-d, 73.23.Ad, 73.40.-c} 

\narrowtext

\section{Introduction}
   
Electron motion in a perpendicular magnetic field and a spatially
periodic potential is one of the fascinating problems in physics.
Hofstadter discussed the existence of a self-similar band structure as
due to the commensurability of the flux per unit cell in terms of the
flux quantum. \cite{hofs92}  Since this level structure is inaccessible
in real solids, due to the need to have a magnetic field of the order
of $10^{5}$T, Hofstadter suggested studying artificial two-dimensional
lattices with much larger lattice spacing than in natural crystals and
corresponding lower fields.   In the last few years it has become
possible to build high-mobility heterojunctions with lateral surface
superlattices and `antidot' arrays.  The beautiful energy spectrum
Hofstadter predicted (his ``butterfly'') has been pursued
experimentally and theoretically in systems of antidots in
semiconductors with small periods. \cite{gen-revs}  On the other hand,
as the lattice spacing is made much larger than the Fermi wave length,
the electron dynamics reaches a semi-classical regime.  In this limit,
it turns out that a competition between the classical cyclotron radius
and the potential length scale (the lattice period) determine a great
deal of the dynamical behavior, as we will see below.  Given the great
flexibility in system fabrication, it is now possible to study the full
range of this problem experimentally: from the fully quantum regime to
the semiclassical mechanics problem of the dynamics of ballistic
electrons in a spatially-modulated potential in a magnetic field.
\cite{gen-revs,orig,Fleischmann96}

In the semiclassical regime, commensurability oscillations in the
magnetoresistance of modulated two-dimensional electron gases have
attracted much attention recently. \cite{gen-revs}  The
commensurability oscillations result from the competition between two
length scales: the cyclotron radius $R_c = v/\omega _o$ (where $v$ is
the particle velocity,  $\omega _o =eB_o/mc $, and $B_o$ is the applied
magnetic field), and the period of the superstructure $a$.  The case
where the potential barriers are defined by an electrostatic modulation
has been studied intensively both theoretically and experimentally.
\cite{gen-revs,Fleischmann96}  The low field oscillations in the 
magnetoresistance have been observed by several groups and in different 
regimes.  For low and moderate fields (Fermi wavelength $\lambda
_F \ll R_c$), and in high mobility samples (mean free path $\ll a$),
the Landau level quantization can be neglected and a classical approach
for ballistic electrons is indeed sufficient to describe the dynamics
and the magnetotransport.  In this regime, Wagenhuber {\em et al.}
studied theoretically the electron dynamics in a square
electrostatically-generated lattice and showed that the chaotic
behavior is reflected in the low-frequency power spectrum of the
system, giving rise to a peculiar type of anomalous particle diffusion.
\cite{Wagenhuber11}

The maxima in the diagonal elements of the magnetoresistance tensor
$\rho _{ii}$ as a function of magnetic field have been attributed to
the existence of pinned electron trajectories around a single maximum
(or groups of them) in the potential landscape. \cite{gen-revs} This
pinning leads to a reduction in magnetoconductance $\sigma _{ii}$, with
minima at fields such that $2R_c /a =n - 1/4$, where $n$ is an integer.
 At the same time, the so-called runaway orbits represent skipping 
orbits along the rows of the potential landscape, and contribute to
enhance $\sigma _{ii}$.  Fleischman {\em et al}. studied the case for a
square geometry by working the classical dynamics and the
(zero-frequency) DC-transport numerically, \cite{Fleischmann96} and
their results where in excellent agreement with experimental values.
\cite{gen-revs}  Schuster {\em et al.} showed experimentally that the
asymmetry of rectangular antidot superlattices is reflected in the
measured DC-transport, as the scattering with the antidots make the
motion clearly more diffusive in one direction than the other.
\cite{Rchuster12}

In this paper we investigate a model where the second length scale in
the problem, apart from the cyclotron radius $R_c$, is defined through
a periodic variation in the magnetic field itself, instead of an
additional electrostatic modulation.  This magnetically modulated
system  was introduced by Vasilopoulos and Peeters, \cite{Peeters92} and
successfully implemented by several groups. \cite{Ye90}  In this
case, it has been  shown that for a weak modulation similar minima in
$\sigma_{ii}$ occur here but are shifted to $2R_c /a = n + 1/4$.
\cite{Peeters92}  This phase shift has in fact been used to differentiate
between these two effects in experiments, \cite{Ye90} and has been
shown to persist beyond the weak modulation regime.
\cite{Peeters92,Gerhardts}

For the square geometry in the magnetic modulated system, we have shown
in an earlier paper that the chaotic orbits are the ones that
contribute the most to the DC-conductivity, \cite{TokyoUS} and
reproduced the experimental results for the commensurability
oscillations in the DC-magnetoresistivity (similar results have been
obtained by Schmidt \cite{Schmidt13}).   In the (finite frequency)
AC-transport we showed also that the quasiperiodic (pinned) orbits give
rise to resonance peaks at characteristic frequencies.  In the magnetic
square lattice the chaotic contribution to the AC-conductivity is
centered around $\omega _o$, while the quasiperiodic trajectories give
rise to features at the frequencies associated with the  rate of
precession of their orbits and/or other characteristic frequencies of
the motion. \cite{TokyoUS}  Similar qualitative behavior has been
reported in experiments by Vasiliadou {\em et al}. in an electrostatic
square antidot array. \cite{Vasiliadou94}  They mapped experimentally
the photoconductivity signals {\em vs}. the uniform field, and found a
clear resonant signal related to quasiperiodic orbits around groups of
(or single) antidots.  These experiments are performed at low magnetic
field and with frequencies in the microwave regime.  Vasiliadou {\em et
al.} found that the commensurability effects and modified classical
cyclotron resonances they observed are in agreement with model
calculations based on the nonlinear dynamics and classical transport of
the electron. \cite{Fleischmann96,Vasiliadou94}  

For the more general
magnetic {\em rectangular} modulation, one would expect that the
quasiperiodic orbits would be also reflected in the
frequency-dependent magnetotransport, and reflect the system anisotropy.
 There are in fact three types of trajectories in this geometry: pinned
and runaway {\em quasiperiodic}, as well as a type of runaway {\em
chaotic}, as we will show below.  This classification refers to their
spatial behavior as the dynamics progresses, and it has been used to
intuitively understand their contribution to the conductivity. 
Although the pinned quasiperiodic and runaway chaotic orbits exist in
the square geometry, \cite{TokyoUS} the runaway quasiperiodic
trajectory is possible only in the asymmetric modulation of a
rectangular geometry, and in a regime of parameters such that the
classical cyclotron orbit radius is comparable to the modulation
periods.  This highlights another interesting point in these systems. 
We show here that the electron dynamics depends only on the ratios of
cyclotron orbit radius to lattice periods, $R_c /a$, and $R_c /b$, and
to the ratio of magnetic modulation to uniform field component, $r=B_m
/ B_o$.  We can therefore say that the chaotic character of the
dynamics is controlled not only by the modulation amplitude, as one
would expect, but by the size of the cyclotron orbit radius.  This is
somewhat different to the near-independence on cyclotron radius
described by Wagenhuber {\em et al}. in the square lattice
electrostatic case. \cite{Wagenhuber11} 

We also find, as will be discussed in sections III and IV below, that
the chaotic orbits in the rectangular lattice differ markedly from the
square lattice case, depending on the value of the ratios $R_c /a$ and
$R_c /b$.  This leads to contributions to the AC-conductivity much
different than in the case of the square lattice, as unstable
quasiperiodic orbits embedded in the dynamics dominate $\sigma _{ii}
(\omega)$ even in a fully chaotic regime.  Finally, we show in this
paper that there is a direct correlation between the largest Liapunov
exponent of the chaotic trajectory and $\sigma _{ii}$.  The Liapunov
exponent provides a direct measure of the diffusion of the particle in
the chaotic orbits, even overwhelming the remnant impurity scattering
(which otherwise provides a low-frequency cutoff in the power spectrum
and yields normal particle diffusion).

\section {Model and approach}

Consider a 2DEG in the $xy$ plane with a spatially modulated magnetic
field, giving rise to a smooth and infinitely extended `magnetic
antidot potential'.  Perhaps the simplest form of this modulation can
be described by the following expression (easily generalizable by
adding more Fourier components, for example),
 \begin{equation} 
 {\bf{B}} =\hat{z} B_o\left[1+\frac{r}{2}\left(\cos\frac{2\pi x}{a}
 + \cos\frac{2\pi y}{b}\right)\right],
 \end{equation}
 where $r= {B_m}/{B_o}$ is the ratio of modulation and uniform field
components ($r \leq 1$), and $a$ and $b$ are the periods of modulation
along the two directions.  This system can be described by the
Hamiltonian $H = \left[{ \bf{p}} + \frac {e}{c}
{\bf{A}}\left({\bf{r}}\right)\right]^2 /2m$, where $m$ is the
electronic effective mass (in GaAs, for example, the system of choice
in typical experiments)  and $\bf{A}$ is the vector potential. 
Choosing a symmetric gauge, this can be written  as,
 \begin {equation}
 {\bf{A}} = \left(-B_o\frac{y}{2}-\frac{B_m}{2k_y}\sin k_y y \,
 , \, B_o\frac{x}{2}+\frac {B_m}{2k_x}\sin k_x x \, , \, 0 \right),
 \end{equation}
 where $k_x =2\pi /a$, and $k_y =2\pi /b$. The trigonometric functions
in the Hamiltonian lead to a nonlinear coupling in the classical
equations of motion.  This coupling is proportional to the
dimensionless quantity $r$, and no analytical solution can be given in
general. We study the dynamics of the system by carrying out accurate
numerical solutions of the equations of motion, and use Liapunov
exponents to characterize the type of trajectory and their contribution
to the conductivity of the system.  Although the resulting equations
can be solved in a linearized form for small $r$, this solution loses
validity rather quickly as $r$ increases.  Only the numerical solutions
are presented here.

In calculating the AC-conductivity we use a classical version of the
linear response theory (Kubo formula)
 \begin {equation}
 \sigma_{\rm ij} (\omega) \propto \int_0^\infty
 dt~e^{-t/\tau}~e^{i\omega t}~\left<V_{\rm i}(t) V_{\rm j}(0)\right>,
 \label{sigmaeq}
 \end {equation}
 where $\left< V_{\rm i} (t) V_{\rm j} (0) \right> = N_{\rm ic}^{-1}
\sum_{{\rm ic}} V_{\rm i} (t) V_{\rm j} (0) $ plays the role of the
velocity auto-correlation function, and the characteristic ensemble
average has been substituted by an average over initial conditions (ic)
in this four-dimensional phase space $\{ \vec {r}_o, \vec {v}_o \}$.
(Since we are only interested in the frequency dependence and magnetic
field features of the conductivity, we ignore an overall normalization
prefactor in \ref{sigmaeq}.) Here, $N_{ic}$ is the total number of
initial conditions used, and $\tau $ is a phenomenological scattering
time associated with the remnant random impurity and alloy scattering
in the real system. \cite{tau-note}  We calculate the velocity
correlation function $\left<V_{\rm i}(t) V_{\rm j}(0)\right>$ by
generating random sets of initial conditions $\{ \vec {r}_o, \vec {v}_o
\}$, while keeping the energy constant.  We can then separate the
initial conditions that yield pinned and runaway orbits by using the
Poincare surface section to classify trajectories for such energy. 
This is accomplished by introducing a ``diffusion length'', $d=[(x_f
-x_o)^2 +(y_f -y_o)^2]^{1/2}$, where $(x_o,y_o)$ is the initial
position, and $(x_f,y_f)$ is the position at the end of the integration
at time $t_f$ ($\gg \omega_o^{-1}$).  The orbit is classified as
runaway if $d > qR_c$, for $q >5$, say.  We use these initial
conditions to generate the Poincare surface section anew and verify
whether the trajectory is indeed runaway-chaotic or
runaway-quasiperiodic, for example.  By identifying the pinned {\em
vs}. runaway trajectories, we can then directly and quantitatively
correlate their characteristic frequencies with their contribution to
the total transport coefficients, as we describe below.  Clearly, the
precise cutoff value of $q$ above would affect results, but only very
slightly quantitatively, and not our conclusions.

Notice that in this Hamiltonian system the kinetic energy is a constant
of motion, since only magnetic fields are applied.  One can define the
cyclotron radius $R_c = v / \omega_o = \sqrt{\dot{x}^2+\dot{y}^2}/
\omega _o$ as an auxiliary length scale.  Using this, the classical
equations of motion can be scaled in time and length by $x/R_c
\rightarrow \tilde{x}$,  $y/R_c \rightarrow \tilde{y}$, and $ \omega_o
t \rightarrow \tilde{t}$.  The equations then appear as those for two
nonlinear coupled pendula,
 \begin{eqnarray}
 \ddot{\tilde{x}} &=& - \dot{\tilde{y}} \left [ 1 + \frac{r}{2} \left(
 \cos k_a \tilde{x} + \cos k_b \tilde{y} \right) \right ] \nonumber \\ 
 \ddot{\tilde{y}} &=& + \dot{\tilde{x}} \left [ 1 + \frac{r}{2} \left(
 \cos k_a \tilde{x} + \cos k_b \tilde{y} \right) \right ] \, ,
 \label{scaled}
 \end{eqnarray}  
 where $\dot{\tilde{x}} = d \tilde{x} / d \tilde{t}$, etc., $k_a = 2
\pi R_c /a$, and $k_b = 2\pi R_c /b$.  This scaling shows the explicit
dependence of the motion on only three parameters: $r$ (magnetic field
modulation amplitude), $R_c/a$, and $R_c/b$, which then fully
characterize the dynamics, as we will show in the next sections.

\section {Poincare surfaces of section and Liapunov exponents}
 
 Since energy is conserved in this system, the dynamics takes place on
a three-dimensional `slice' of the phase space available.  We can fully
characterize the types of trajectories by means of Poincare surfaces of
section on the $xy$ plane for a given value of velocity (or `phase'),
such as $\dot{x}$ maximum (and $\dot{y} = 0$).  Due to the
translational symmetry of the superstructure potential in the $xy$
plane, we fold all Poincare sections such that the $x$ values lie in
the interval $(-a/2 , a/2)$, and the $y$ in $(-b/2, b/2)$.  The phase
space trajectories are uniquely determined by points in the surface of
section.

 We calculate Liapunov exponents by using the method of Wolfe.
\cite{Wolf}  We choose base 2 in calculating exponents, such that the
distance between two nearby trajectories is $d(t) = d_0 2^{\lambda _{i}
t}$, where $\lambda _i$ are the Liapunov exponents.  For a Hamiltonian
system, $\sum_i \lambda _i =0$, since the volume in phase space is
conserved  by Liouville's theorem. \cite{liouville}  For a given set of
parameters $r$, $R_c/a$ and $R_c/b$, the type of trajectory depends
only on the initial conditions $\left( \vec{r}_o, \vec{v}_o \right)$,
as described above.
 
For a certain modulation strength $r$, we find that the fractions of
runaway chaotic, runaway quasiperiodic, and pinned quasiperiodic orbits
in phase space depend only on the length ratios $R_c/a$, and $R_c/b$,
but not on the energy nor the uniform field individually. This
dependence, consequence of the scaling shown in Eq.\ (\ref{scaled}), is
reflected on the AC-conductivity as well.  For example, in the case of
a square lattice case, a Poincare surface section and three different
traces of $\sigma _{xx}(\omega)$ are shown in Fig.\ 1, where $r=0.6$
and $2R_c/a=1.4$ are kept constant.  These traces were produced
changing the energy such that $v = 0.7 \omega_o a$, while the
corresponding uniform field is given by $\omega_o \tau= 1.5, 2.25$, and
3.0 ($\tau = 3 \times 10^{-12}$ sec).  In this situation, we find that
the Poincare sections do not change, and the traces of conductivity
have the same features, except for a frequency shift towards $\omega
_o$, and a different amplitude.  The horizontal shift in Fig.\ 1(b) is
nothing but the scaling of the frequency (or time) by $\omega_o$ (or
$\omega_o^{-1}$).  The amplitude change can be understood if one
analyzes the expression for the conductivity in a {\em uniform} field,
giving the classical Drude peak with half-width $1/\tau$,
 \begin{equation}
 \sigma _{xx}(\omega) \propto  \frac {\tau v^2}{1 +(\omega -
 \omega_o )^2\tau ^2} \, .
 \label{Eq6}
 \end{equation}
 It is clear from this expression that the conductivity amplitude
depends on the energy (via $v^2$).  In our case of a modulated field,
$\sigma_{xx} (\omega)$ shows more structure, and it appears at the
expense of the Drude peak.  Rescaling of the frequency to $\omega /
\omega_o$ in each trace produces the nesting one would expect from the
scaling of the equations (4), as clearly shown in the inset of Fig.\
1(b).  As we will explore further in the next section, the various
features observed in these traces are contributions from the different
quasiperiodic regions in the Poincare map, while the contribution of
the chaotic region is centered close to $\omega _o$. \cite{TokyoUS}

As explained, the characteristic size of the cyclotron orbits with
respect to the magnetic lattice dimensions determines the resulting
dynamics and the overall properties of the trajectories for different
initial conditions.  We can then consider different regimes: {\em
First}, when $2R_c \leq a \leq b $, most orbits are localized between
or around modulation maxima (single `antidots'), as the high magnetic
field (and/or low energy) effectively shrinks the electronic orbits to
be fully within a period of the modulation.  {\em Second}, when $ a
\leq 2R_c \leq b $, the asymmetry of the potential landscape is
expected to be strongly reflected in the dynamics and transport, since
all length scales are comparable and their competition produces strong
changes in the dynamics.  {\em Third}, for low magnetic fields (and/or
high energies), when $a \leq b \leq 2R_c$, the lattice asymmetry
becomes less and less important, as the trajectories extend over
several periods of the potential landscape.  Here, the particle motion
effectively performs a self-averaging of the different magnetic field
amplitudes, which cancels the asymmetry of the system, and yields a
relatively large DC-conductance and a featureless and broad
frequency-dependent $\sigma(\omega)$.

 For the {\em first regime}, consider $2R_c/a = 3/8$ and $b/a=1,2$.
Since $2R_c$ is much smaller than $a$ and $b$, the electron is able to
trace out periodic and quasiperiodic orbits which remain basically
pinned about a lattice position, even for a high modulation $r$. These
orbits show a variety of frequencies that can be seen in the structure
of the Poincare section.  For example, when $b/a=1$, $r=0.6$, the
Poincare section is mostly dominated by large sectors of quasiperiodic
orbits, even though the nonlinear coupling (via $r$) is somewhat high. 
We show in Fig.\ 2(a) that two different regions of pinned orbits exist
in this portrait, surrounded by Kolmogorov-Arnold-Moser (KAM) islands
of stability, \cite{liouville} and a small chaotic region showing
diffusion along both directions.  For the chaotic orbit in this case we
calculate the largest Liapunov exponent to be $\lambda = 0.85$ (while
for the quasiperiodic orbits $\lambda _i \approx 0$ --- this is the same
for all the cases below, as anticipated, so that we will only quote the
largest Liapunov exponent in the chaotic orbit from here on).  The
quasiperiodic structures in the Poincare map, as we will see later,
will give rise to resonance peaks in the AC-conductivity.

Figure 2(b) shows a Poincare section where $b=2a$ and $r=0.6$. In this
case, phase space has more structure due to the asymmetry of the
potential landscape, KAM islands are still well developed, and the
chaotic region is larger. In this case, the largest Liapunov exponent
is $\lambda = 0.91$, indicating that the motion in the chaotic regime
has become more diffusive (and still two-dimensional).

In the {\em second regime}, we consider as examples $2R_c/a = 1.2$, and
$b/a \geq 3$ (smaller values of $b/a$ yield results quite similar to
the square geometry discussed before \cite{TokyoUS}).  Since $2R_c$ is
between both $a$ and $b$, there is a preferred direction to the
electron motion, in this case the $x$ direction.  We have shown in
previous work that in the square geometry, when $2R_c \approx a = b$,
the electron dynamics shows a great deal of chaotic behavior {\em even
when the modulation constant $r$ is small}. \cite{TokyoUS}  This can be
understood as being due to the fact that the electron is more likely to
`collide'  in this regime with the maxima in the modulation landscape. 
These `collisions' allow the particle to access stronger nonlinear
terms in the equation of motion, which in turn produces a more critical
dependence to initial conditions and then chaotic dynamics.
\cite{comment}

 For $b \gtrsim 3$, as we turn-on the modulation, we notice the
appearance of two kinds of quasiperiodic orbits (pinned and runaway),
even for $r \lesssim 0.35$.  The runaway quasiperiodic orbits result
from the strong $x$-$y$ asymmetry of the potential, which forces the
electron to move preferentially along the $x$-direction (smaller
period).  However, the modulation field is not sufficiently strong to
make the electron motion chaotic, and the runaway orbits are nearly
free in the preferred direction.  As $r$ increases to $\approx 0.4$,
the chaotic orbits start to occupy a non-negligible volume in phase
space, and the largest Liapunov exponent becomes positive.  In Fig.\
\ref{fig4} we show Poincare sections for $2R_c/a=1.2$, and $b/a=4$, for
both $r=0.45$ and $r=0.85$.  In \ref{fig4}(a), $r$ smaller, we see
three types of trajectories: two quasiperiodic (one pinned and one
runaway or open), and a chaotic runaway.  Notice that the chaotic
trajectories are confined between two sets of quasiperiodic runaway
orbits and yield then particle diffusion {\em only along one
dimension.}  In this case, the chaotic trajectories have a single
characteristic frequency around $\omega_o$, with largest Liapunov
exponent $\lambda = 0.98$.  In Fig.\ \ref{fig4}(b), as $r$ is nearly
doubled, we see that there are two small regions of pinned
quasiperiodic orbits with Liapunov exponents converging to zero
individually, while the rest of phase space is filled by a chaotic
orbit with largest Liapunov exponent $\lambda = 1.1$, which clearly
shows diffusion in two dimensions.  The modulation is so strong that
there are no remnants of the runaway quasiperiodic trajectories.  This
$y$-axis `delocalization' is similar to the energy dependence described
by Wagenhuber {\em et al.} \cite{Wagenhuber11}  for a symmetric
electrostatic modulation case.   Notice here, moreover, that energy is
kept constant, via $2R_c/a$, and that the transition is produced then
by the stronger modulation amplitude.

We find furthermore that the chaotic trajectories are qualitatively
different in this second regime from those in the first high-field
regime. In the latter, when $2R_c \leq a \leq b$, the characteristic
frequency of the chaotic trajectory is always near $\omega_o$, as the
trajectory effectively samples a significant range of field values
during its two-dimensional diffusion and self-averages to $\omega _o$.
In the second regime, however, we find that it is typical to find {\em
two} characteristic frequencies embedded in the chaotic orbit at
$\omega _{\pm}\approx \omega _o \pm \delta$, where $\delta$ increases
as either $b/a$ or $r$ increase. This type of remnant two-frequency
trajectory is produced by the asymmetry of the potential, and one can
clearly see this effect in the special cases where $2R_c \approx a \ll
b$, for example.  These dimensions create a landscape where the
electron can move for a relatively long time in the $x$ direction in a
region of fields lower than $B_o$ (in essence a diffusive motion of its
precession center), corresponding to a frequency $\omega _{-} \approx
\omega _o - \delta$.  When the motion drifts in the $y$ direction, the
electron faces a wall of maxima separated by a distance $a \approx 2R_c
$, which often confine the trajectory to a region of higher fields
between two maxima and pin the particle to precess for a time with
characteristic frequency motion of $\omega _{+} \approx \omega _o +
\delta$, until it escapes again to a lower field region in the
potential.  The persistence of these characteristic frequencies
embedded in the chaotic motion provides a unique $\sigma _{ii} (\omega)
$, as we will see in the next section.

Another form of exploring this intermittence effect is by analyzing the
field experienced by the particle along its chaotic trajectory.  In
Fig.\ \ref{fig3} we show the `instantaneous frequency', $f={e B({\rm\bf
r})}/ {mc}$, which the electron experiences in the magnetic potential
landscape as it moves through positions {\bf r}$(t)$. When $a=b$,
\ref{fig3}(a) shows that this instantaneous frequency $f$ varies around
$\omega _o$ ($= 1.25 \times 10^{12}$ Hz here) nearly uniformly.   In
the second plot, \ref{fig3}(b), the particle's instantaneous frequency
spends a significant amount of time at values averaging clearly higher
than $\omega _o$, and then, in the next segment of the time series, its
average is lower than $\omega _o$. This persistent intermittent
switching behavior, which continues for as long as we have run the
simulations, indicates that the particle is metastably trapped in the
different regions of the periodic potential, even though the trajectory
is clearly chaotic, as judged by the corresponding Poincare phase
portrait.  We will show in the next section that this kind of
trajectory will give rise to two prominent peaks in the
AC-conductivity, even though its Poincare map shows a typical chaotic
trajectory, and one would naively expect a featureless $\sigma
(\omega)$ (which is the case for $2R_c/a \gg a, b$, for example).
Therefore, the peculiar metastable character of the nearly trapped
sections embedded in this chaotic trajectory shows clearly in the
frequency dependent conductivity.  

Finally, for the low-field {\em third regime}, we consider $2R_c/a =
6$, $b/a=2$, and $r=0.56$.  Since $2R_c \gg  a, b$, the electron is
more likely to `collide' with potential maxima here than in the
previous two cases, as discussed earlier, and hence chaotic orbits
occupy most of phase space even for extremely low modulation (not
shown).  In all these cases, the chaotic orbits have an average
characteristic frequency at $\omega \approx \omega _o$, but with a
rather broad distribution, as we will see below.  Moreover, the largest
Liapunov exponent for the chaotic orbits is $\lambda=1.14$ , the
largest exponent we ever obtained in all cases, indicating that the
motion is indeed fully diffusive in two dimensions. This behavior will
be clearly reflected on the magnetotransport described in the next
section.

\section {Magnetotransport}

A quantitative theory of magnetotransport requires detailed
consideration of regular and/or quasiperiodic (pinned or runaway) and
chaotic (runaway) trajectories arising from modulation scattering.  In
Hamiltonian systems, where the volume in phase space is conserved, the
conductivity tensor $ \sigma $ is obtained from the sum of the
individual contributions of trajectories weighted by their volume in
phase space. We have shown in previous work that the contribution of
pinned orbits to the DC-conductivity is negligibly small, in agreement
with previous work, \cite{Fleischmann96} while they are closely related
to peaks in the AC-conductivity.  We assume here that the proportion of
pinned to runaway orbits does not change qualitatively even after the
inclusion of impurity scattering, which should not be far from being
the case if mobility in the unmodulated system is high. \cite{tau-note}
 
For the calculations of the AC-conductivity we used random sets of
initial conditions (typically a few thousands), and classified the
orbits by inspecting the corresponding Poincare sections, as described
above.  We are then able to separate the initial conditions that give
rise to pinned or chaotic orbits, and clearly identify their different
contributions to $\sigma$ and $\rho$. When the modulation strength $r$
is zero, the AC-conductivity yields the classical Drude peak, as the
charge experiences a uniform field and only impurity scattering,
included here through $\tau$ (see Eq.\ (\ref{Eq6})).  Once the magnetic
modulation is turned on, however, the Drude peak remains centered
around $\omega = \omega _o $, for small $r$, but the conductivity
begins to acquire a non-homogeneous broadening, and $\sigma _{xx}
(\omega)$ and $\sigma _{yy} (\omega)$ start deviating from each other
if $b \neq a$, in general.  This `inhomogeneous' broadening appears
{\em before} the chaotic orbits appear ($r$ small), i.e., when the
Poincare sections show only quasiperiodic orbits.  This broadening is
then clearly associated with the fact that most of the electron
trajectories have still the characteristic frequency $\omega _o$. 
However, due to the small modulation, the electron orbits precess with
characteristic frequencies which appear close to or far from $\omega
_o$, depending on structural parameters.   By increasing the modulation
strength, additional features appear in $\sigma (\omega)$ at the same
time that chaos and KAM islands appear in the Poincare surface section.
 Moreover, an offset at zero frequency appears in both $\sigma_{xx}(0)$
and $\sigma_{yy}(0)$ due to the onset of chaotic orbits and associated
particle diffusion.  An increasing portion of chaotic trajectories as
$r$ increases makes the zero-offset increase at the expense of the
resonance peaks.  This signals the increase in DC conductance and drop
in resistance produced by the magnetic modulation which has been
measured in experiments on these and related systems.
\cite{gen-revs,Ye90}  We should also mention that the DC
magnetoresistance calculated here shows the well-known Weiss
oscillations seen recently for magnetic modulation, \cite{Ye90} as we
have shown in Ref.\ [\ref{TokyoUS}].

In this section, we discuss the magnetoconductivity for the same three
regimes of $R_c/a$, $R_c/b$ and $r$ parameter values reviewed in the
previous section.  We discuss the calculated conductivity tensor
components, and their relation to features of the corresponding
Poincare section for those parameter values.  For the {\em first
regime}, where $2R_c < a \leq b$, we have used the same initial
conditions and parameters we used to generate the Poincare surface of
section in Fig.\ 2(a).  Figure 5 shows the AC conductivity for both
$b=a$ and $b=2a$.  In 5(a), since $a=b$, we obviously find $\sigma
_{xx} (\omega)=\sigma _{yy} (\omega)$. The $\sigma _{xx} (\omega)$
curve has two main features around $\omega _o$, related to the two
regions of quasiperiodic trajectories in the Poincare surface of
section in Fig.\ 2(a) which dominate the phase space.  The central
feature with frequency $\omega \approx \omega _o$ is enhanced by the
contribution of the small chaotic region which provides frequencies
closely peaked at around $\omega _o$.  Since $2R_c /a= 3/8$ here, we
see that the effect of the modulation on $\sigma (\omega)$ is
relatively minor, even for the not so small $r=0.6$, producing particle
precession around the minima and maxima of the modulation landscape, in
addition to some higher frequency trajectories associated with pinning
near maxima.

In Fig.\ 5(b) we use the same parameter set used to generate the
Poincare section in Fig.\ 2(b). Since here $a \neq b$, the symmetry is
broken and it is the case that $\sigma _{xx} (\omega) \neq \sigma
_{yy}(\omega)$.  By comparing the Poincare sections 2(a) and 2(b) we
see that most of the quasiperiodic orbits in the central section
survive, which in Fig.\ 5(b) produce the persistence of the feature in
$\sigma _{xx} (\omega)$ and $\sigma _{yy} (\omega)$ at $\omega \approx
1.4 \omega _o$.  On the other hand, the other sector of quasiperiodic
orbits does shrink and becomes surrounded by KAM islands in 2(b). 
Correspondingly, we see that the conductivity feature at $\omega
\approx 0.8 \omega _o$ becomes smaller, due to the shrinking of phase
space volume occupied by these trajectories, and moves to lower
frequencies ($\approx 0.6 \omega_o$).  Meanwhile, the contribution of
the KAM islands in 2(b) gives rise to large amplitudes at various
frequencies $\omega > \omega _o$.   Finally, in this regime we see that
as the symmetry breaks, the zero frequency offset value of $\sigma
_{xx}$ becomes larger than that of $\sigma _{yy}$, indicating that the
electron motion becomes more diffusive in the $x$-direction than the
$y$-direction.  This is expected, as the $b=2a$ geometry produces open
`channels' along the $x$-direction which facilitate diffusion, even in
this first regime.

For the {\em second regime}, take $2R_c/a=1.2$, $b/a > 2$, and
$r=0.85$.  We will focus on the contribution of the chaotic orbits
discussed in the previous section, in connection with Fig.\ 4.  In this
range of parameters, most of phase space volume is occupied by chaotic
trajectories.  In Fig.\ 6(a), we show several traces of $\sigma_{xx}
(\omega)$, keeping $2R_c=1.2a$ and $r$ constants, while changing $b/a$
as indicated there. For $1 < b/a \le 2$, we should mention that there
is no appreciable difference from the case of a square geometry,
probably due to the fact that as $2R_c$ is comparable to the lattice
parameters $a$ and $b$, the electron scatters more frequently and
symmetrically, producing chaotic orbits with characteristic frequency
self-averaging to $\omega _o$. When $ b/a$ grows, the asymmetry in the
lattice structure increases, and $b$ is less comparable to $2R_c$, so
that the electron dynamics changes substantially, as shown in Fig.\
6(a).  We see here that large $b/a$ values produce two prominent
features in $\sigma_{xx}$ at $\omega \approx \omega_o \pm \delta$,
which become increasingly apart in frequency for larger $b/a$.  The
existence of characteristic frequencies different from $\omega_o$, {\em
even when the trajectories are fully chaotic} (see Fig.\ 3(b)), are due
to the long-term metastability of the nearly-pinned orbits discussed in
relation with Fig.\ 4(b).  It is then interesting to see that even
though the Poincare section shows a chaotic map, $\sigma (\omega)$ is
able to exhibit this unusual metastable behavior arising from (and made
stronger by) the asymmetry in the magnetic modulation landscape.  We
should stress that even in the case of a rather short $\tau$ ($= 3
\times 10^{-12}$ sec in this figure), the metastability is not `erased'
by the impurity scattering, and is clearly visible in $\sigma_{xx}$.

In Fig.\ 6(b), we keep $2R_c/a=1.2$ and $b=4a$, while changing $r$ as
indicated in the graph (in this figure, and to contrast with 6(a), we
use $\tau = 15 \times 10^{-12}$ sec, three times longer than before). 
For weak modulation, $r<0.5$, the runaway quasiperiodic trajectories
and KAM-tori do survive and the chaotic trajectories do not cross the
wall of maxima in the $y$-direction, as shown earlier in Fig.\ 3(a). 
Moreover, the corresponding chaotic trajectory extended along only the
$x$-direction will self-average to a frequency $\omega \approx \omega
_o$.  This can clearly be seen in $\sigma_{xx} (\omega)$ curves which
show only a broadening similar to a Drude peak for a uniform field
(except for a larger width and flatter top produced by the intrinsic
diffusive behavior of the trajectories). As the modulation increases to
$r \ge 0.5$, we start seeing the single $\sigma$-peak split into a
double peak about $\omega _o$. At $r=0.85$, where most of the phase
space is occupied by a chaotic trajectory, the double peak is clearly
developed (in addition to a low-frequency remnant structure from the
small quasi-periodic orbits). We should also notice that in this regime
the zero frequency offset of $\sigma _{xx}$ is always larger than that
of $\sigma _{yy}$ (not shown), in agreement with the intuitive notion
that the runaway orbits along the $x$-direction contribute more to the
DC-conductivity, even when the diffusion is fully two-dimensional.

In order to further identify the different contributions to the
conductivity in this regime, we now calculate the spectrum for a
single-orbit, $S_{xx} (\omega)$. This quantity is the contribution to
$\sigma_{xx}$ coming from a single trajectory (i.e., the velocity
autocorrelation function for that trajectory).  Figure 7 shows $S_{xx}$
for the {\em same} chaotic orbit but starting at two different points
on the trajectory. The first trace, appearing solid and with a peak at
$\omega \approx 0.9 \omega_o$, is obtained when we integrate the
dynamics starting at a point in the magnetic landscape identified as a
`channel' of minima, i.e., the electron motion is mostly diffusive
along the $x$-direction (an integration over a $5\tau$ interval is
presented).   The second curve, dashed and with a peak at $\omega
\approx 1.3 \omega _o$, was obtained when the integration starts at a
point between two nearest antidots (maxima) in the $x$-direction.
Notice that here the electron motion was drifting to cross the magnetic
barrier in the $y$-direction and formed a short-lived metastable
quasiperiodic orbit.  It is then further verification that as the
particle executes the chaotic trajectory, it is being trapped along
`channels' formed by nearest neighbor maxima or antidots.  This effect
is carried through to the frequency-dependent conductivity, even for
the short $\tau$ used ($= 3 \times 10^{-12}$ sec).

For the {\em third regime} of large cyclotron orbits, let us consider
$2R_c/a =6$, $b/a=2$, and $r=0.56$.  As mentioned in the previous
section, the phase space in all these cases is completely occupied by a
single chaotic trajectory, even for weak magnetic modulation strength 
$r$, and shows not signs of metastability.  One typically finds $\lambda
=1.14$, the largest Liapunov exponent we found for this regime.  We show 
in Fig.\ 8 the conductivity curves of $\sigma _{xx}
(\omega )$ and $\sigma _{xy} (\omega )$ for the case mentioned. In this
example, even though $a \neq b$, we obtain $\sigma _{xx} (\omega )=
\sigma _{yy} (\omega )$ due to the strong scattering produced by the
modulation, since on the scale of the cyclotron radius the lattice
appears basically symmetric.  Consequently, the chaotic trajectory
averages over all directions, producing  a single but much broadened
peak at $\omega = \omega_o$ and with a large zero-frequency offset,
showing indeed that the motion is fully diffusive in both directions.

 \section {Conclusions}

 We have studied the frequency-dependent magnetotransport in a 2D
magnetic field modulation in a rectangular lattice symmetry for various
parameter regimes.  This study has revealed that a class of resonances
exist in the AC-conductivity, reflecting the different character of the
various electron trajectories, and the degree of integrability (or
non-) of these systems.  In all cases, we have found a correlation
between the DC conductivity $\sigma _{ii}(0)$ and the value of the
largest positive Liapunov exponent $\lambda $. As $\lambda $ increases,
the zero-frequency offset increases, indicating that the chaotic
trajectories become more and more diffusive, even changing diffusive
character from one- to two-dimensional, as the modulation increases. 
The study of different profiles of magnetic field modulation and even
different lattice structures, both theoretically and experimentally,
should give us better insights into  the microscopic character of the
electron trajectories in different regimes.  The possible screening
effects which would mask some of the frequency dependence discussed
here (magnetoplasmon effects) are being studied and will be presented
elsewhere. \cite{Heitmann,Ulloa10}  It is however anticipated that
the single-particle features discussed here would persist even when
plasmon effects are taken into consideration, as the experiments in
Ref.\ [\ref{Vasiliadou94}] have shown.

\acknowledgements

We thank helpful discussions with J. Thomas, R. Rollins and P. Jung.  
This work has been partially supported by DOE Grant No.\
DE-F02-91ER45334.  SEU acknowledges support of the  AvH Foundation.

\begin{figure}[htb]
 \caption{(a) Poincare surface of section for square modulation with 
 $2R_c /a = 2v/\omega_o=1.4$, and $r=0.6$. (b) $\sigma _{xx}(\omega)$ 
 for the same parameters but different energy ($v$) and magnetic field 
 ($\omega_o$).  Frequency $\omega$ in units of $10^{12}$ Hz.  Inset: 
 traces rescaled to $\omega / \omega_o$ show perfect nesting. } 
 \vskip0.1cm 
 \label{fig1}
\end{figure}

\begin{figure}[htb]
 \caption {Poincare surfaces of section for the same energy, with
 $2R_c /a=3/8$, and $r=0.6$. In (a) $b=a$, while in (b) $b=2a$.} 
 \vskip0.1cm
 \label{fig2}
\end{figure}

\begin{figure}[htb]
 \caption { Poincare surfaces of section for the same energy,
 and $2R_c/a=1.2$, and $b=4a$.  For (a) $r=0.45$; and (b) $r=0.85$.
 Notice $x$-direction one-dimensional diffusion of chaotic orbits in
 (a) evolves into two-dimensional diffusive motion in (b). } 
 \vskip0.1cm
 \label{fig4}
\end{figure}

\begin{figure}[htb]
 \caption {Time series of the `instantaneous frequency'  $f= 
 {eB({\bf r})}/{mc}$ that particle experiences in a chaotic trajectory.
 Here, $\omega_o = 1.25$ ($\times 10^{12}$ Hz), while $2R_c/a=1.2$ and
 $r=0.85$.  In (a) $a=b$, while in (b) $b=4a$.  Time axis is labeled by
 the integration step index.} 
 \vskip0.1cm 
 \label{fig3}
\end{figure}

\begin{figure}[htb]
 \caption {Frequency-dependent conductivities, $\sigma (\omega)$.
 Parameters in (a) are, $2R_c /a=0.4$, $b/a=1$, and $r=0.6$, and
 $\sigma_{xx} = \sigma_{yy}$.  In (b) same parameters,
 except for $b/a=2$.  Here, solid (dashed) line shows $\sigma_{xx}$
 ($\sigma_{yy}$).}
 \vskip0.1cm 
 \label{fig5}
\end{figure}

\begin{figure}[htb]
 \caption {(a) $\sigma_{xx} (\omega) $ for $2R_c /a=1.2$, $r=0.85$,
 $\tau = 3 \times 10^{-12}$ sec, and $b/a = 2, 3, 4, 8$, as shown.  (b)
 $\sigma_{xx} (\omega)$ shown for $2R_c/a = 1.2, \, b/a=4$, $\tau$
 three times longer, and $r=0.25, 0.5$, and 0.85, as shown.} 
 \vskip0.1cm
 \label{fig6}
\end{figure} 

\begin{figure}[htb]
 \caption {Single-orbit spectrum $S_{xx} (\omega)$ for the same chaotic
 trajectory with $2R_c /a=1.2$, $b/a=4$, and $r=0.85$, but starting at
 different points.  Solid trace reflects motion along $x$-channels;
 dashed trace that between nearest antidots.  }
 \vskip0.1cm 
 \label{fig7}
\end{figure}

\begin{figure}[htb]
 \caption {$\sigma_{xx} (\omega) $ (solid trace) and $\sigma _{xy}
 (\omega) $ (dashed) for $2R_c =6a$, $b=2a$, and $r=0.56$. $xx$ and $yy$
 components are identical.} 
 \vskip0.1cm 
 \label{fig8}
\end{figure}

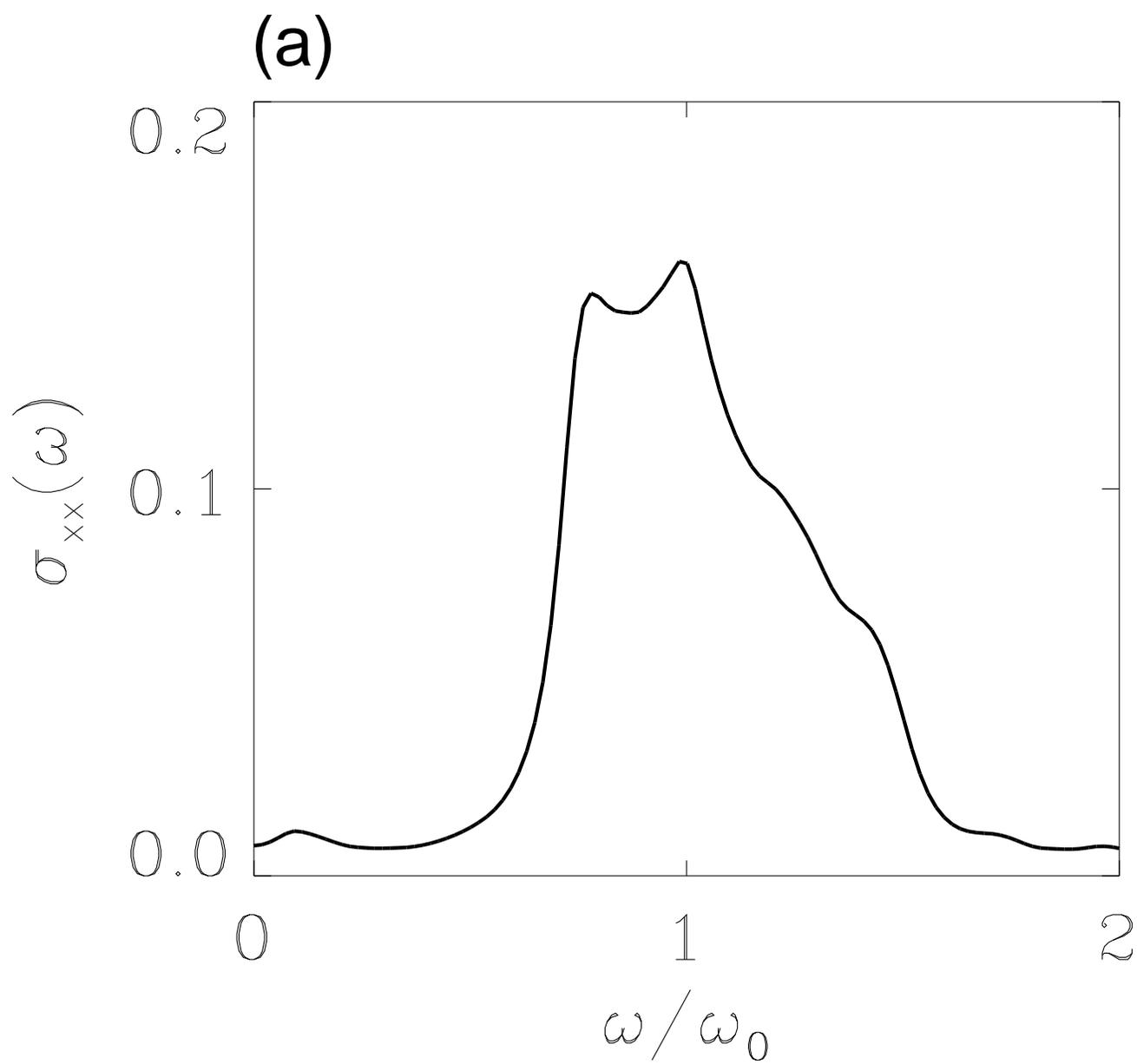
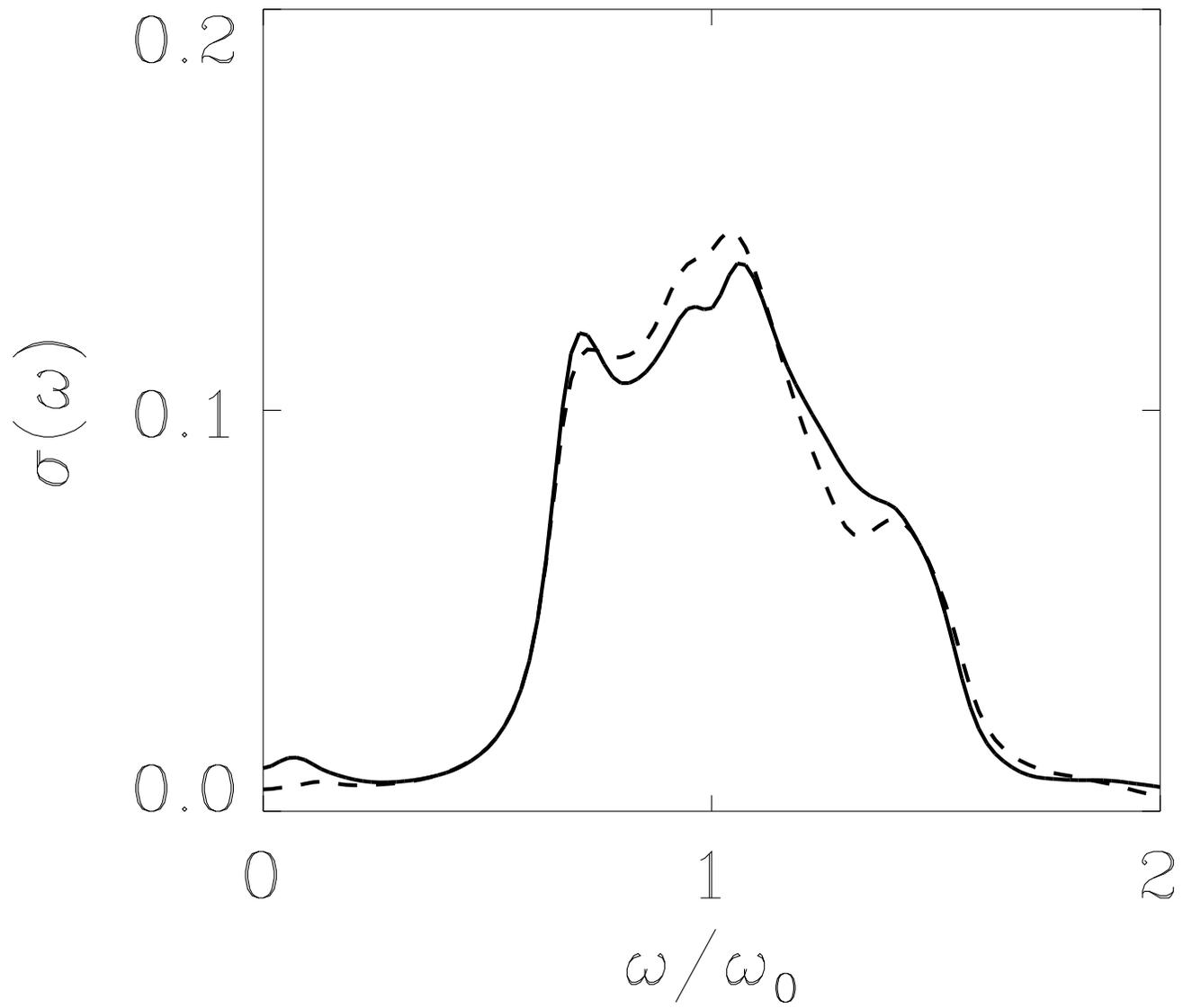
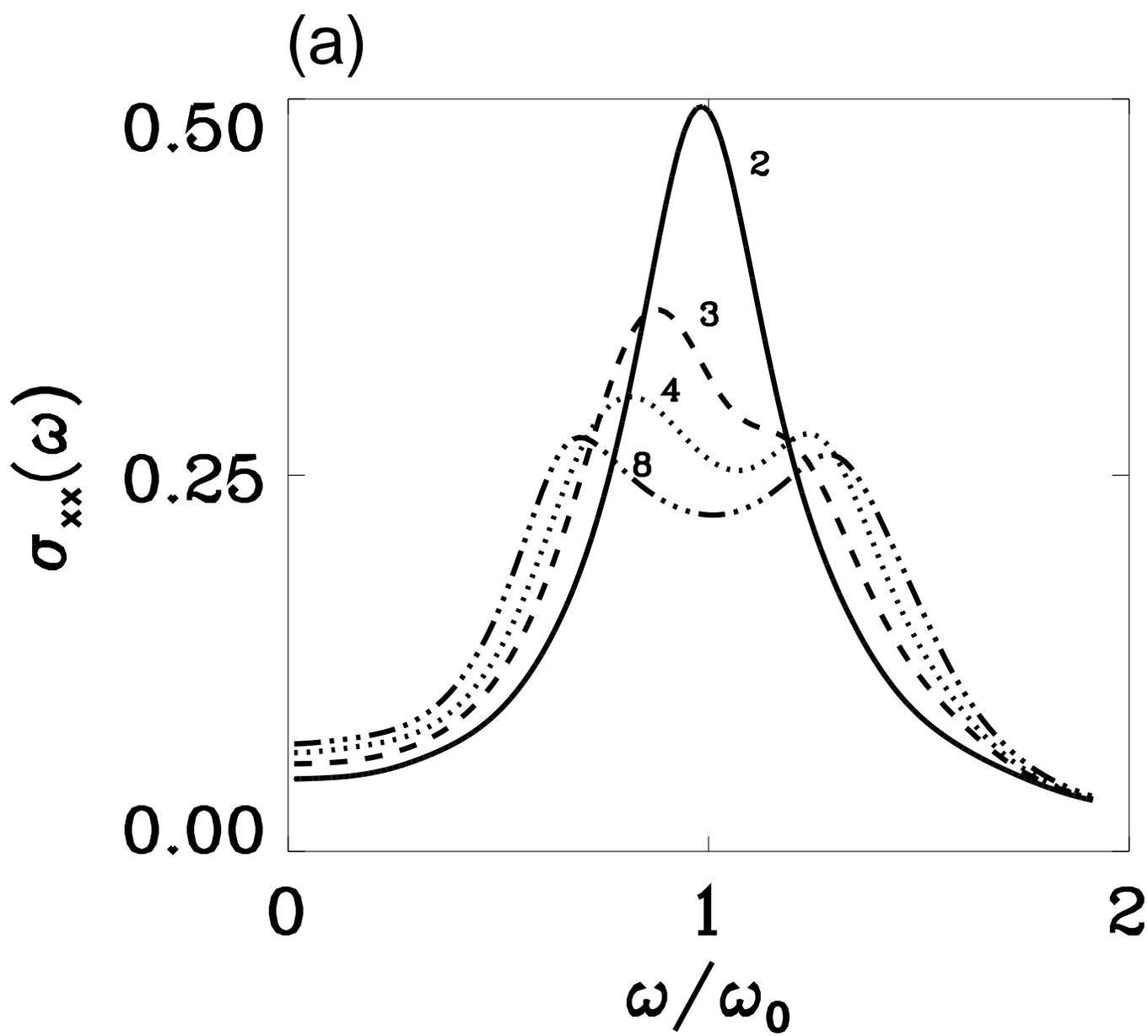
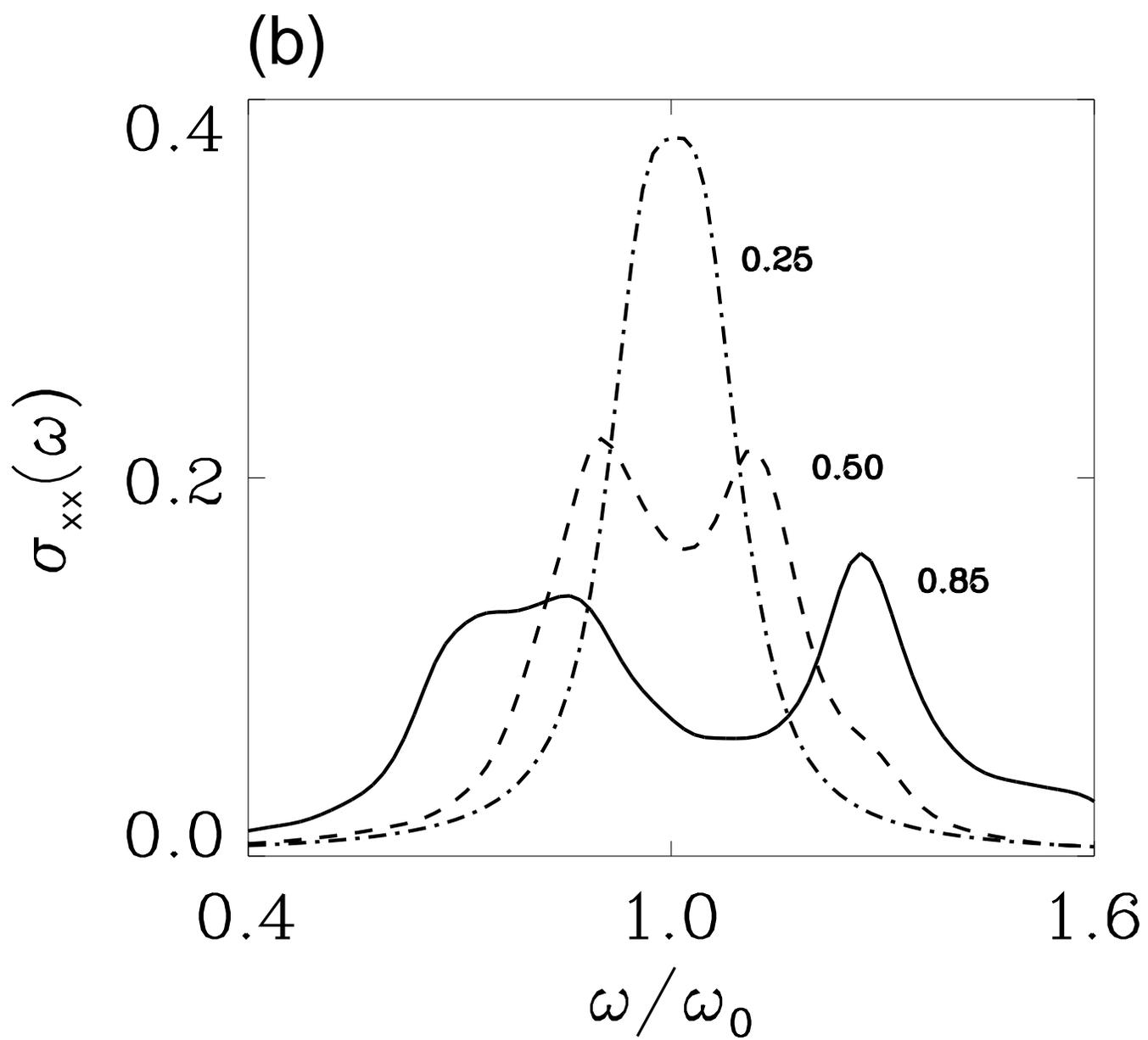
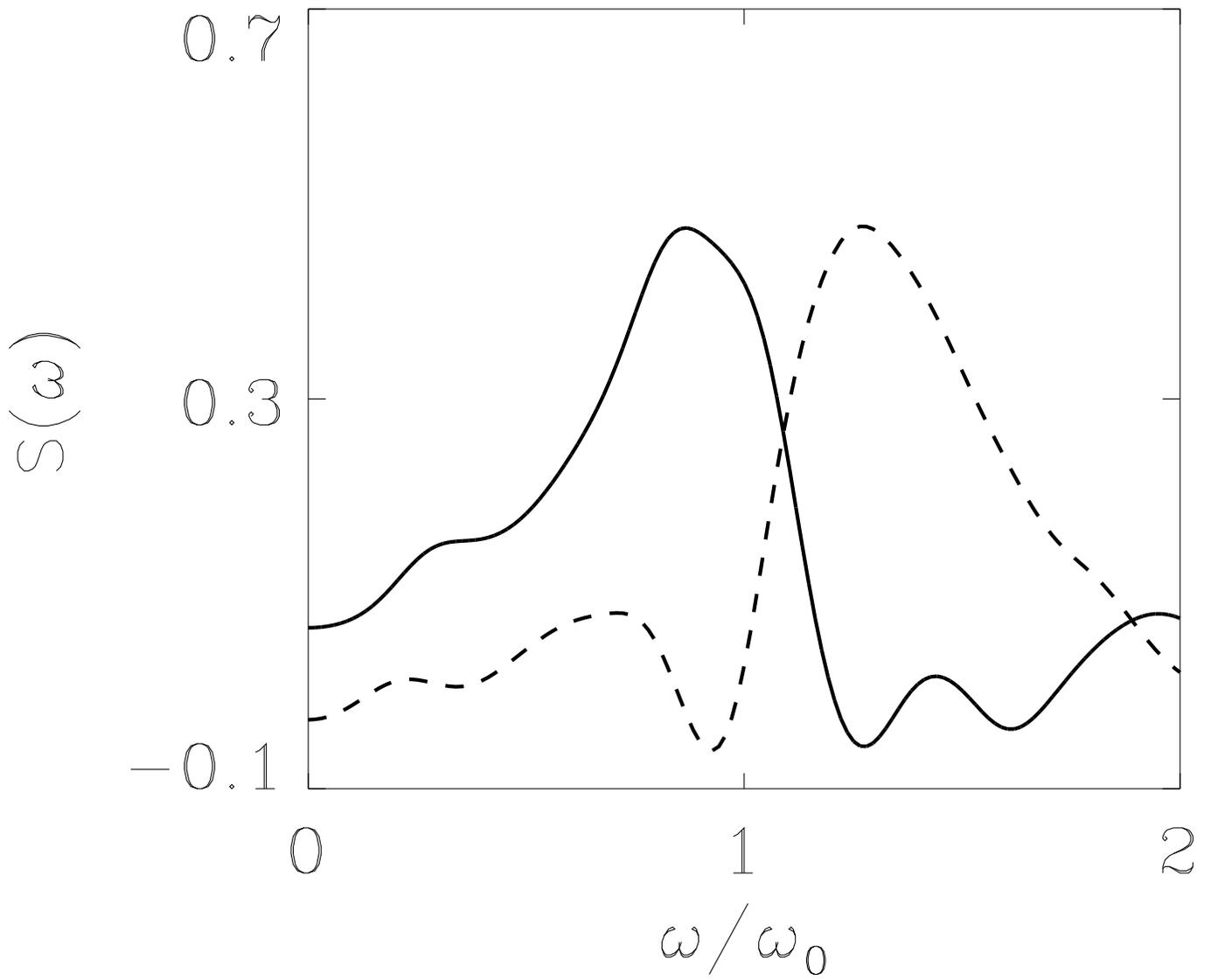
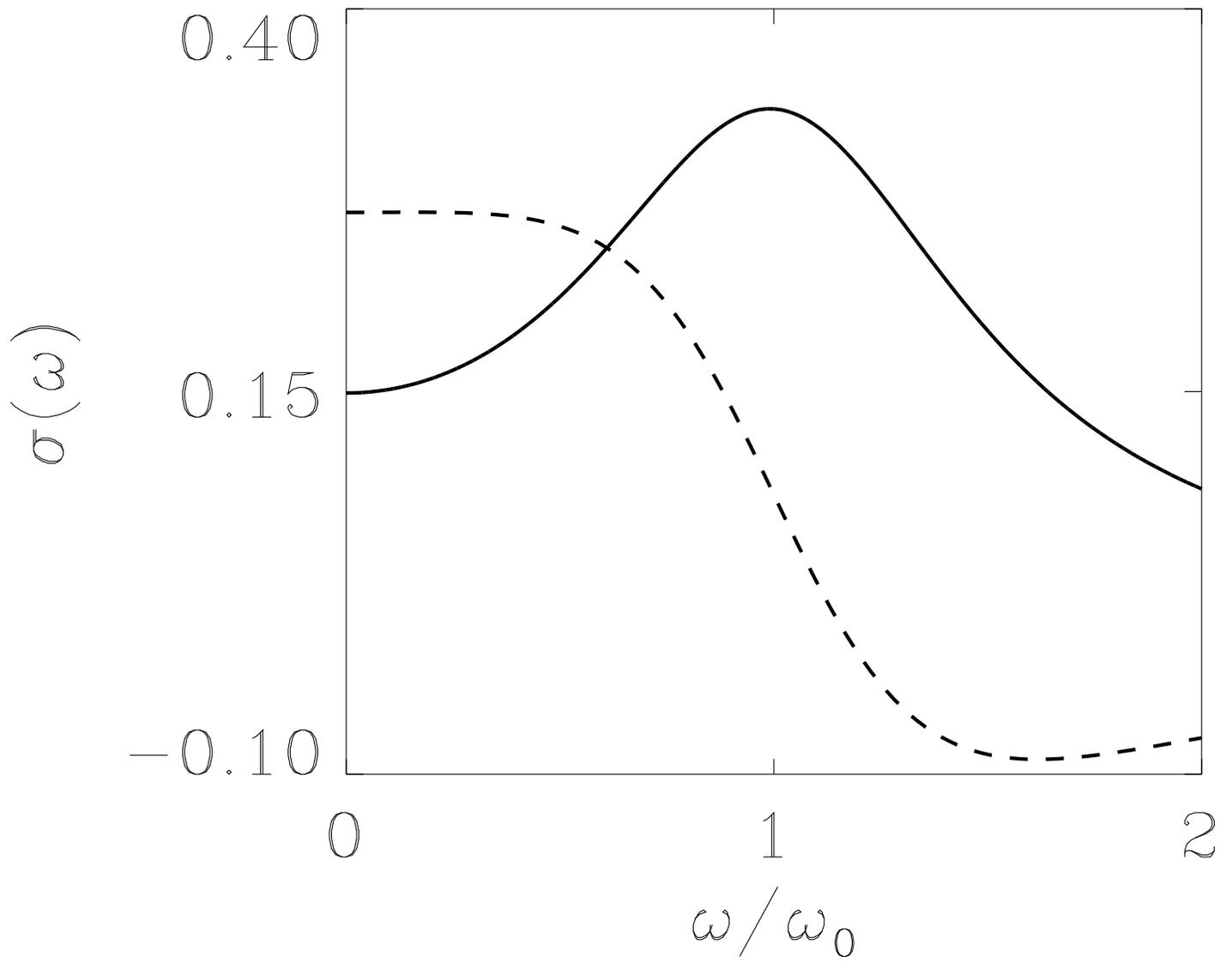
\end{document}